\documentclass[10pt,leqno]{amsart}
\usepackage{graphicx}
\usepackage{indentfirst,csquotes}
\usepackage{blindtext}

\usepackage{soul,color}
\usepackage{amssymb,amsthm,amsmath}
\usepackage{xcolor,paralist,hyperref,titlesec,fancyhdr,etoolbox}
\usepackage{tabularx}

\titleformat{\section}[display]{\normalfont\huge\bfseries\centering}{\centering\chaptertitlename\thechapter}{10pt}{\Large}
\titlespacing*{\section}{0pt}{0ex}{0ex}

\hypersetup{ colorlinks=true, linkcolor=black, filecolor=black, urlcolor=black }

\usepackage{lipsum}

\begin{document}
\title{Influence of Spatial Coherence on Phonon Transmission across Aperiodically Arranged Interfaces}
\author{Theodore Maranets*}
\author{Milad Nasiri}
\author{Yan Wang}
\address{Department of Mechanical Engineering, University of Nevada, Reno, Reno, NV, 89557, USA}
\email{tmaranets@unr.edu}
\maketitle

\begin{abstract}
In this study, we employ the atomistic wave-packet method to directly simulate coherent phonon transport and scattering dynamics in an aperiodic superlattice structure with aperiodically arranged interfaces. Our investigation reveals that interference dynamics, contingent upon the relative magnitudes of phonon wavelength, spatial coherence length, and average interface spacing, profoundly influence transmission spectra and mode-conversion behavior. Particularly, longer-wavelength phonons possessing larger coherence lengths exhibit more pronounced destructive interference and mode-conversion, leading to generally lower transmission. The insights gained into coherent phonon wave physics from this study can significantly augment the analysis and design of phononic devices.

\end{abstract} 

\bigskip

The study of phonon transport, scattering, and localization within metamaterials characterized by secondary periodicity, such as superlattices (SLs) and nanomeshes, has garnered substantial research attention \cite{zhang2021coherent,anufriev2021review}. The purposeful manipulation of these processes holds substantial promise for advancing a diverse range of applications. In SLs, for instance, interfaces can efficiently scatter phonons, and with optimization of the SL structure, achieving ultralow lattice thermal conductivity ($\kappa_{L}$) that approach or even break the random-alloy-limit $\kappa_{L}$ \cite{yao1987thermal,venkatasubramanian2000lattice,landry2008complex,wang2015optimization}. Consequently, SLs have garnered attention as potential thermoelectric materials due to their ultralow $\kappa_{L}$ \cite{venkatasubramanian2000lattice,snyder2008complex}. On the other hand, it is crucial to explore strategies for enhancing heat dissipation in devices incorporating SL-like structures, such as transistor arrays and quantum cascade lasers.

It has been theorized that thermal transport in systems featuring a secondary periodicity is dictated by phonon coherence, particularly at low temperatures. Notably, in SLs, the emergence of a minimum $\kappa_{L}$ as the period length decreases to a few atomic layers is attributed to a transition from interface scattering to coherent wave transport \cite{venkatasubramanian2000lattice,simkin2000minimum,chen2005minimum,ravichandran2014crossover}. The existence of two transport regimes is explained by the presence of two distinct types of phonon modes \cite{simkin2000minimum,yang2003partially,wang2014decomposition}. The first type represents phonon modes belonging to the constituent base materials of the SL, e.g., the phonon modes of bulk silicon (Si) and germanium (Ge) within a Si/Ge SL. The second type includes phonon modes aligning with the secondary periodicity of the SL, following the dispersion relations of the SL rather than those of the base materials. The former is referred to as incoherent phonons, while the latter is denoted as coherent phonons, a terminology adopted in this letter. These two phonon types induce different length dependencies (i.e., the number of periods) of $\kappa_{L}$ in SLs \cite{wang2014decomposition,luckyanova2012coherent}. When coherent phonons dominate, $\kappa_{L}$ can almost linearly increase with the number of SL periods due to the ballistic transport of coherent phonons. Conversely, if incoherent phonons dominate, $\kappa_{L}$ remains nearly constant, constrained by interface scattering effects. Moreover, the two phonon types impact the temperature dependence of $\kappa_{L}$ \cite{chakraborty2020complex} and interfacial transmission is affected by the nonequilibrium of coherent and incoherent phonons \cite{ma2022ex}.

Recent advancements have been made in understanding coherent phonon transport in SLs, specifically, the conceptual introduction of spatial coherence \cite{latour2014microscopic,latour2017distinguishing}. Spatial coherence length $l_{c}$, defined as the real-space width of a phonon wave-packet, serves as a measure of the spatial correlations of atomic motion within the phonon wave. $l_{c}$ has been demonstrated to influence the interference states, thereby effectively characterizing the transition from incoherent to coherent transport in SLs \cite{latour2014microscopic}. Due to the relative infancy of this theory, there is a present scarcity of research \cite{wang2014decomposition,luckyanova2012coherent,lee2017investigation,hu2018randomness,juntunen2019anderson,hu2020machine} elucidating the propagation, scattering, and localization of incoherent and coherent modes in SLs or nanomeshes, particularly those in aperiodic forms, within the context of spatial coherence. A rigorous and quantitative analysis of these behaviors is imperative for a comprehensive understanding of their unique characteristics, which is pivotal for developing effective strategies to tailor the $\kappa_{L}$ of SL(-like) structures for diverse applications, either enhancing or reducing it as needed.

In this letter, we conduct, to our knowledge, the first application of the atomistic phonon wave-packet method to study coherent phonon scattering in a modified SL structure in contrast to previous works that have solely explored incoherent phonon transport in SLs \cite{schelling2003multiscale,jiang2021total}. As demonstrated with the method by Latour and Chalopin \cite{latour2017distinguishing}, coherent phonons travel through the periodic SL perfectly with zero interface scattering. Coherent phonons are scattered, however, when the periodic structure is modified, such as the case of aperiodic layering \cite{tamura1990acoustic,bliokh2009transport}. The character of coherent phonons in aperiodic SLs has mostly been indirectly analyzed through macroscopic coefficients such as $\kappa_{L}$ and/or thermal boundary conductance in the existing literature \cite{wang2015optimization,wang2014decomposition,chakraborty2020complex,juntunen2019anderson,hu2020machine,roy2022unexpected}. Leveraging the wave-packet method, it becomes feasible to directly simulate the transport of phonons with well-defined polarization and wavelength $\lambda$ \cite{schelling2003multiscale,schelling2002phonon}. Furthermore, we can uniquely specify the value of  $l_{c}$ along with the size $L$ of the aperiodic structure. Harnessing the method's unique strength in controlling these three degrees of freedom, we investigate the variations in transmission spectra of coherent phonons with spatial coherence, ultimately shedding light on the wave dynamics within the aperiodic SL structure that impacts heat conduction.

\begin{figure}
\includegraphics[width=\textwidth]{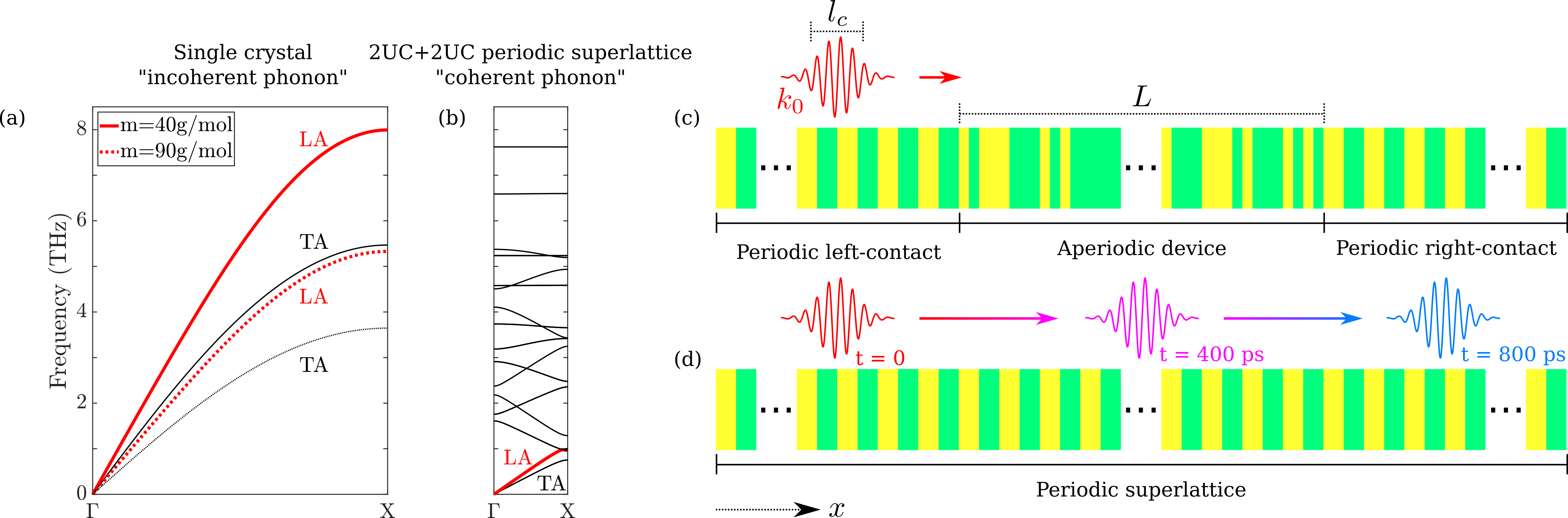}
\caption{Phonon dispersion relations along the cross-plane or [100] direction for incoherent phonons (a) and coherent phonons
(b) of the material system employed in this work with the longitudinal-acoustic (LA) and transverse-acoustic (TA) branches indicated. (c) Schematic illustration of the simulation domain. An aperiodic SL device of thickness $L = 250$ nm is sandwiched between two
periodic SL contacts. The incident LA-mode coherent phonon wave-packet of wavevector $k_{0}$ and spatial coherence length $l_{c}$ is generated in the left-contact
and allowed to propagate into the device. The total energies of the left
and right-contacts, as well as the device are monitored throughout
the simulation to compute the transmission per Eq. 4. (d) Schematic illustration of the coherent phonon wave-packet
propagating perfectly without scattering in the periodic SL.
Note: the illustrated wave-packets are not physically accurate as the
minimum wavelength for a coherent phonon is two periods.}
\label{fig:widedispersionschematic}
\end{figure}

The atomistic method applied in this work is adapted from the technique developed by Schelling et al. \cite{schelling2002phonon} We chose a Lennard-Jones (LJ) solid argon structure as our model material system for reduced computational cost. The LJ model for interatomic interactions is given as \cite{jones1924determination}:
\begin{eqnarray}
    \phi_{ij}(r_{ij}) = 4\epsilon\left[\left(\frac{\sigma}{r_{ij}}\right)^{12} - \left(\frac{\sigma}{r_{ij}}\right)^{6}\right]\qquad .
\end{eqnarray}
The molecular weights of the two layered materials are 40 $g/mol$ and 90 $g/mol$. The thickness of each layer in the periodic SL is 2 unit-cells (UC). The parameter set of  $\epsilon_{Ar}=0.0104$ $eV$, $\sigma_{Ar}=0.34$ nm, and a cutoff radius of $2.5\sigma_{Ar}$ has been used in previous studies of LJ-modeled solid argon which possesses van der Waals bonding in nature \cite{landry2008complex,huberman2013disruption}. In this work, we use $\epsilon=16\epsilon_{Ar}$ for the interactions of all atom types to mimic stronger bonded materials such as Si/Ge and GaAs/AlAs as done in previous LJ-based atomistic investigations \cite{wang2015optimization,wang2014decomposition,chakraborty2020complex,chakraborty2020quenching}. The relaxed lattice constant is $a=5.269$ $\text{\AA}$, making the period thickness $\sim2$ nm. The cross-plane incoherent and coherent phonon dispersion relations for this material system are presented in Figs. ~\ref{fig:widedispersionschematic}a and ~\ref{fig:widedispersionschematic}b, respectively. 

The wave-packet simulations are conducted in the following process. First, a wavevector $k_{0}$ is chosen from the longitudinal-acoustic (LA) branch of the cross-plane coherent phonon dispersion in Fig.~\ref{fig:widedispersionschematic}b. The generated wave-packet will thus be centered at $k_{0}$ in reciprocal-space. For this $k_{0}$, there is an associated frequency $\omega_{0}=\omega(k_{0})$ and group velocity $v_{g0} = v_{g}(k_{0})$ from the LA branch. The wave-packet is generated at the beginning of the simulation by applying the following equations for atom displacement and velocity:
\begin{eqnarray}
u_{i,n} = \frac{A_{i}}{\sqrt{m_{i}}}\varepsilon_{k_{0},i}\exp{(i[k_{0}\cdot(x_{n} - x_{0})-\omega_{0}t]})\exp{(-4(x_{n}-x_{0}-v_{g0}t)^{2}/l_{c}^{2})}\qquad .
\label{eqn:pwp}
\end{eqnarray}
\begin{eqnarray}
v_{i,n} = \frac{\partial u_{i,n}}{\partial t}\qquad .
\end{eqnarray}
where the real parts of $u_{i,n}(t=0)$ and $v_{i,n}(t=0)$ are the initial displacement and velocity for the $i$th atom in the $n$th unit cell, respectively. $A_{i}$ is the amplitude of the wave-packet, $m_{i}$ is the mass of atom $i$, $\varepsilon_{k_{0},i}$ is the eigenvector of the $i$th atom, $x_{n}$ is the position of the $n$th unit cell, and $x_{0}$ is the initial position of the wave-packet center. Notably, $l_{c}$ in Eq.~\ref{eqn:pwp} corresponds to the full width at half maximum of the Gaussian-shaped wave-packet. Phonon properties are determined from lattice dynamics calculations. 

Fig.~\ref{fig:widedispersionschematic}c depicts the setup of the simulation domain. An aperiodic SL device of length $L=250$ nm is sandwiched between two periodic SL contacts each of length $4,875$ nm, making the total size of the system 10 $\mu$m. The device is constructed per the methodology described in Ref.~\cite{wang2015optimization}. The cross-section of the simulation domain is $4 \times 4$ UC$^{2}$. Periodic boundary conditions are applied to all three dimensions. The initial wave-packet is generated sufficiently away from the device and the duration of the simulation is such that by the end, nearly all of the incident energy has been either reflected or transmitted through the device. Following the wave-packet initialization, the simulation is performed in a microcanonical ensemble. The LAMMPS software \cite{thompson2022lammps} is used to implement the atomistic simulations.

The transmission spectra $\mathcal{T}(k)$ across the aperiodic SL is computed by
\begin{eqnarray}
    \mathcal{T}=\frac{E_{rc,final}}{E_{lc,initial}}\qquad .
\end{eqnarray}.
where $E_{rc,final}$ is the total energy transmitted into the right-contact (rc) by the end of the simulation and $E_{lc,initial}$ is the total energy in the left-contact (lc) at the beginning of the simulation. If the device were a periodic SL, then the entire domain would be periodic and $\mathcal{T}=1$ as the coherent phonon travels perfectly with zero interface scattering. This scenario is illustrated in Fig.~\ref{fig:widedispersionschematic}d.

\begin{figure}
    \includegraphics[width=0.7\textwidth]{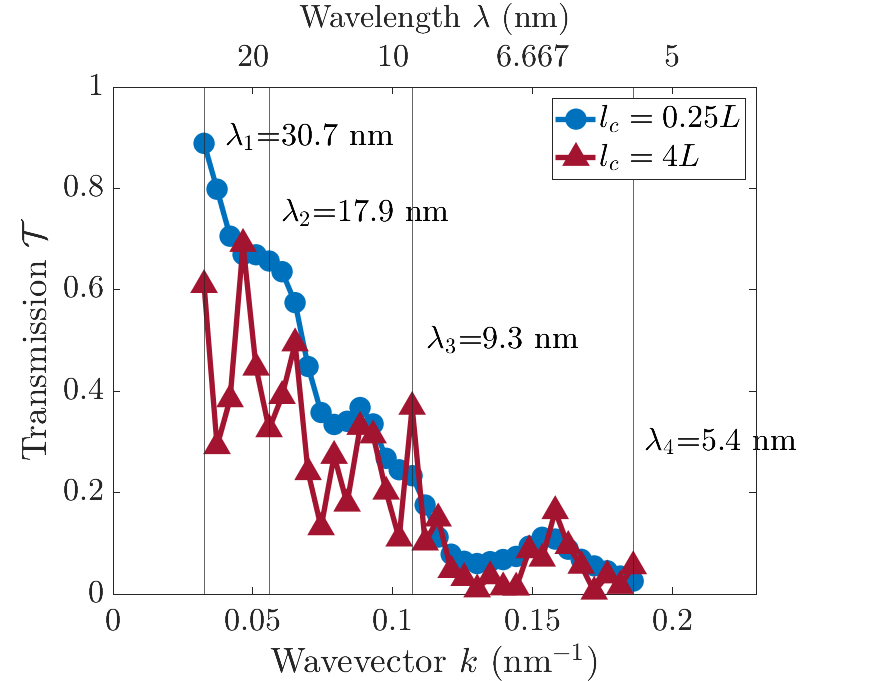}
    \caption{Transmission $\mathcal{T}$ versus wavevector $k$ (or inverse of wavelength $\lambda$) for the LA-mode coherent phonon wave-packet passing through the aperiodic SL device illustrated in Fig.~\ref{fig:widedispersionschematic}c. The spectra curves correspond to wave-packets of different spatial coherence lengths. Key wavelengths $\lambda_{1}=30.7$ nm, $\lambda_{2}=17.9$ nm, $\lambda_{3}=9.3$ nm, and $\lambda_{4}=5.4$ nm analyzed in this letter are indicated in the figure. Very low wavevector data points are omitted as these wavelengths are larger than the small-$l_{c}$ and thus their ballistic transport cannot be simulated.}
    \label{fig:figspectra}
\end{figure}

In Fig.~\ref{fig:figspectra}, we present $\mathcal{T}(k)$ for the LA-mode coherent phonon wave-packet with $l_{c}=0.25L$ and $l_{c}=4L$, representing two scenarios where the spatial coherence length is significantly shorter or longer than the aperiodic SL device length. Evidently, the transmission of the large-$l_{c}$ wave-packet is generally lower than that of the smaller counterpart ($l_{c}=0.25L$) at long wavelengths. As the wavelength decreases, the curves converge, indicating a reduced influence of spatial coherence on the transmission of shorter-wavelength coherent phonons in the aperiodic SL device. 

Remarkably, the observed increase in transmission with decreasing coherence length introduces new perspectives into the understanding of phonon transport in aperiodic SL devices. As previously reported by Chakraborty et al. \cite{chakraborty2020complex}, the thermal conductivity $\kappa_{L}$ of aperiodic SLs increases with temperature, attributed to intensified inelastic phonon transmission across interfaces within the aperiodic SL, akin to the behavior observed in single interfaces \cite{hopkins2009multiple}. Our coherence-length-dependent transmission results in Fig.~\ref{fig:figspectra} unveil a distinct mechanism for the temperature-dependent $\kappa_{L}$ of aperiodic SLs, absent in the case of a single interface. Specifically, Latour et al. \cite{latour2014microscopic} noted a reduction in spatial coherence length at higher temperatures. Accordingly, we attribute the increased $\kappa_{L}$ of aperiodic SLs at elevated temperatures to the shortened coherence lengths at higher temperatures, which increases coherent phonon transmission as demonstrated by Fig. ~\ref{fig:figspectra}. Recognizing the monotonic correlation between temperature and coherence length, our work underscores the potential of the wave-packet method as a robust tool for exploring phonon behaviors, particularly those associated with coherence length, at finite temperatures, despite the simulation being conducted at a background temperature of 0 K.

The influence of spatial coherence on phonon transmission across the aperiodic SL device arises from two major contributing factors, as illustrated in Fig. S1 of Supplemental Materials. Firstly, it governs the spectrum of phonon wavelengths that a wave-packet can accommodate in reciprocal space. Specifically, a wave-packet with larger $l_{c}$ encompasses a narrower spectrum of phonon wavelengths centered around $k_{0}$, thereby increasing the likelihood of robust interference among the phonons within the wave-packet, particularly destructive interference within the aperiodic SL structure.  Secondly, it controls the extent to which a wave-packet can span multiple phonon wave periods in real-space. As $l_{c}/\lambda$ increases, the number of phonon wave periods encompassed by the wave-packet increases proportionally. This creates more opportunities for the transmitted and (multiple-)reflected components to destructively interfere with each other in the aperiodic SL device. 

\begin{table*}
\tiny
\caption{\label{tab:table1}Snapshots of the atom velocity versus real-space position when the wave-packet scatters through the aperiodic SL device for wavelengths $\lambda_{1} = 30.7$ nm, $\lambda_{2} = 17.9$ nm, and $\lambda_{4} = 5.4$ nm at spatial coherence lengths $l_{c} = 0.25L$ and $l_{c} = 4L$, where $L = 250$ nm. Movies of these scenarios are located in the Supplemental Materials. The dotted vertical lines denote the position and thickness of the device.}
\begin{tabular}{|ccc|}
\hline
Fig. 3a: $\lambda_{1}= 30.7$ nm, $l_{c}=0.25L$, $t=460$ ps&Fig. 3b: $\lambda_{2}= 5.4$ nm, $l_{c}=0.25L$, $t=460$ ps&Fig. 3c: $\lambda_{4}= 5.4$ nm, $l_{c}=0.25L$, $t=560$ ps\\ \hline
\includegraphics[width=0.33\textwidth]{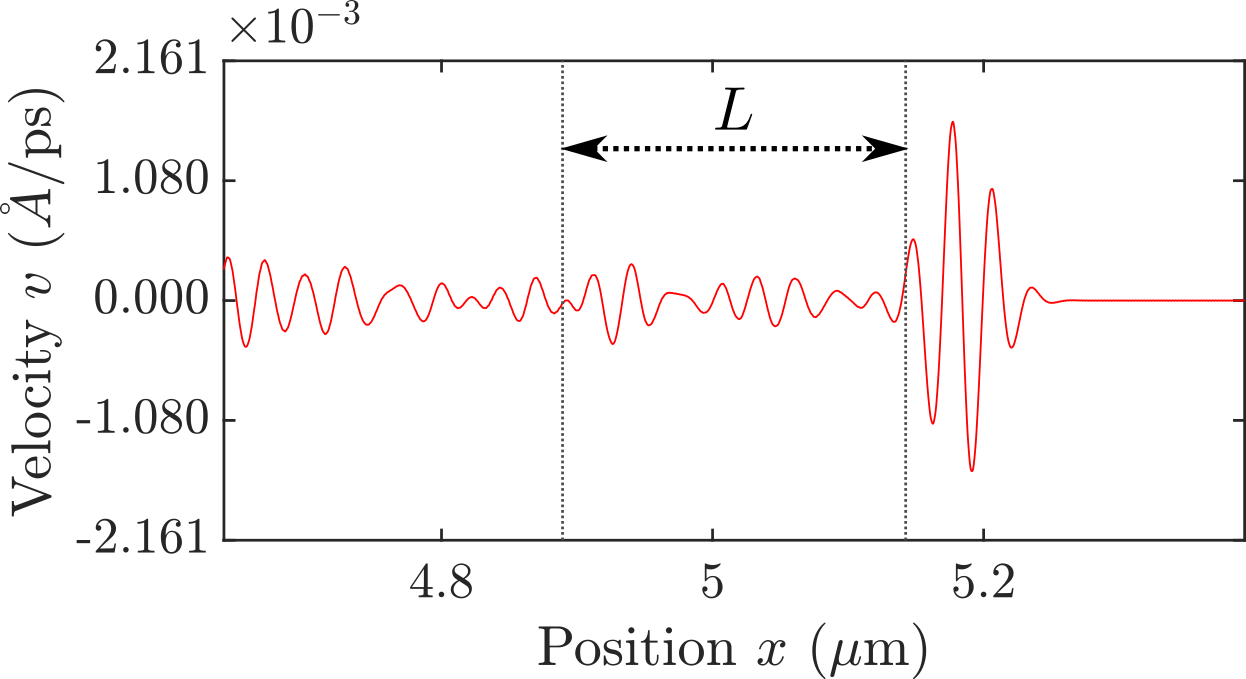}&\includegraphics[width=0.33\textwidth]{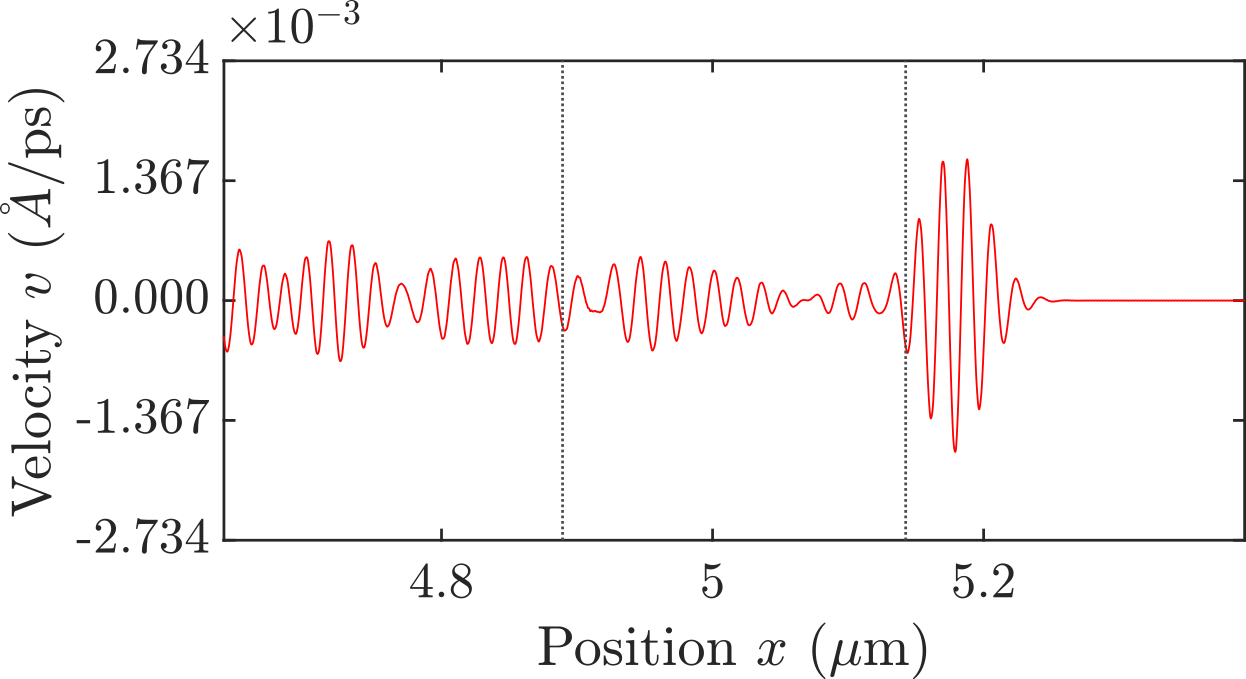}&\includegraphics[width=0.33\textwidth]{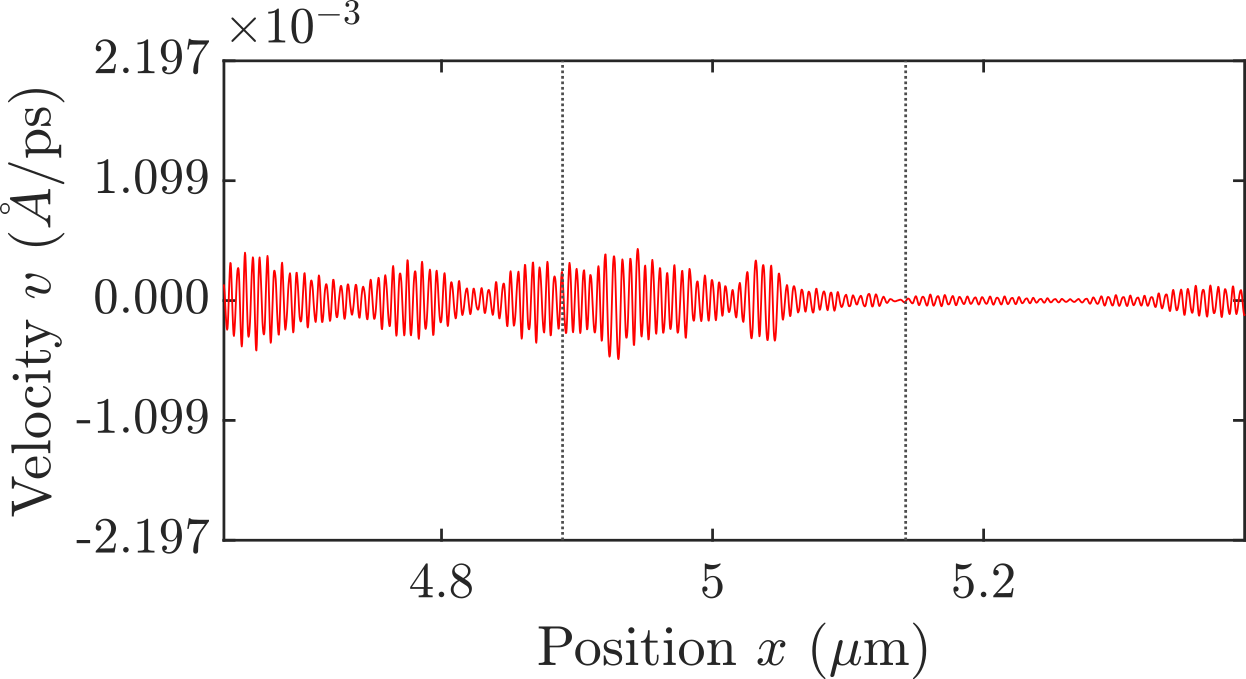} \\ \hline
Fig. 3d: $\lambda_{1}= 30.7$ nm, $l_{c}=4L$, $t=460$ ps&Fig. 3e $\lambda_{2}= 5.4$ nm, $l_{c}=4L$, $t=460$ ps&Fig. 3f: $\lambda_{4}= 5.4$ nm, $l_{c}=4L$, $t=560$ ps\\ \hline
\includegraphics[width=0.33\textwidth]{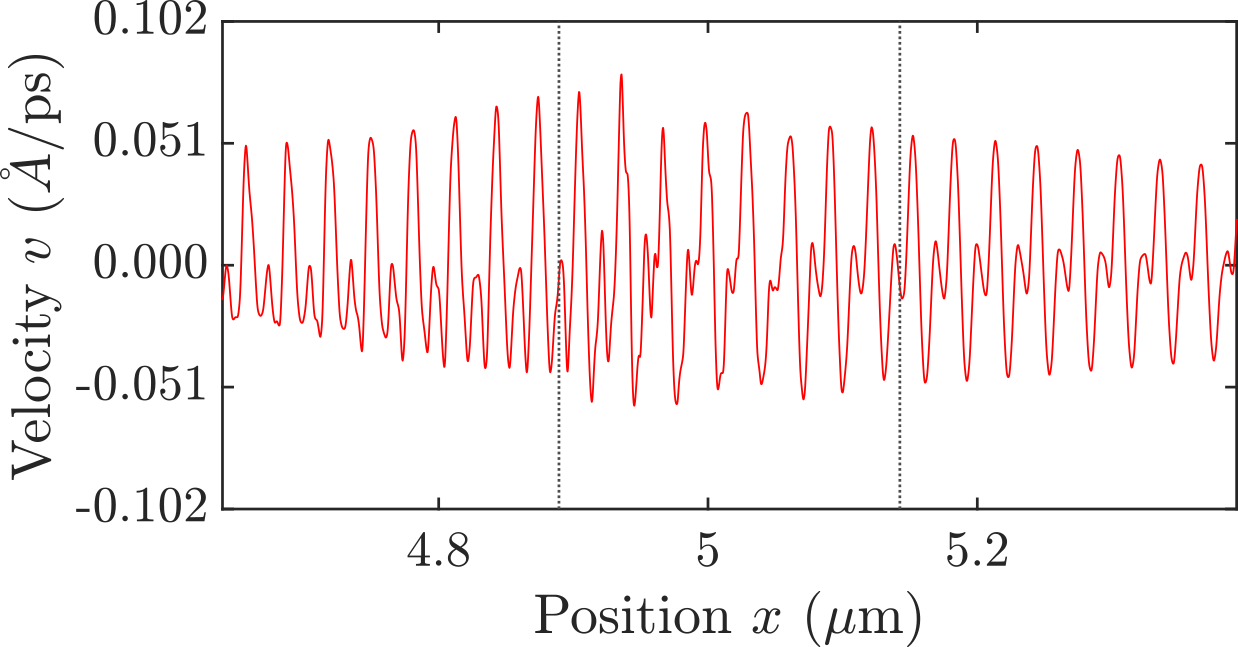}&\includegraphics[width=0.33\textwidth]{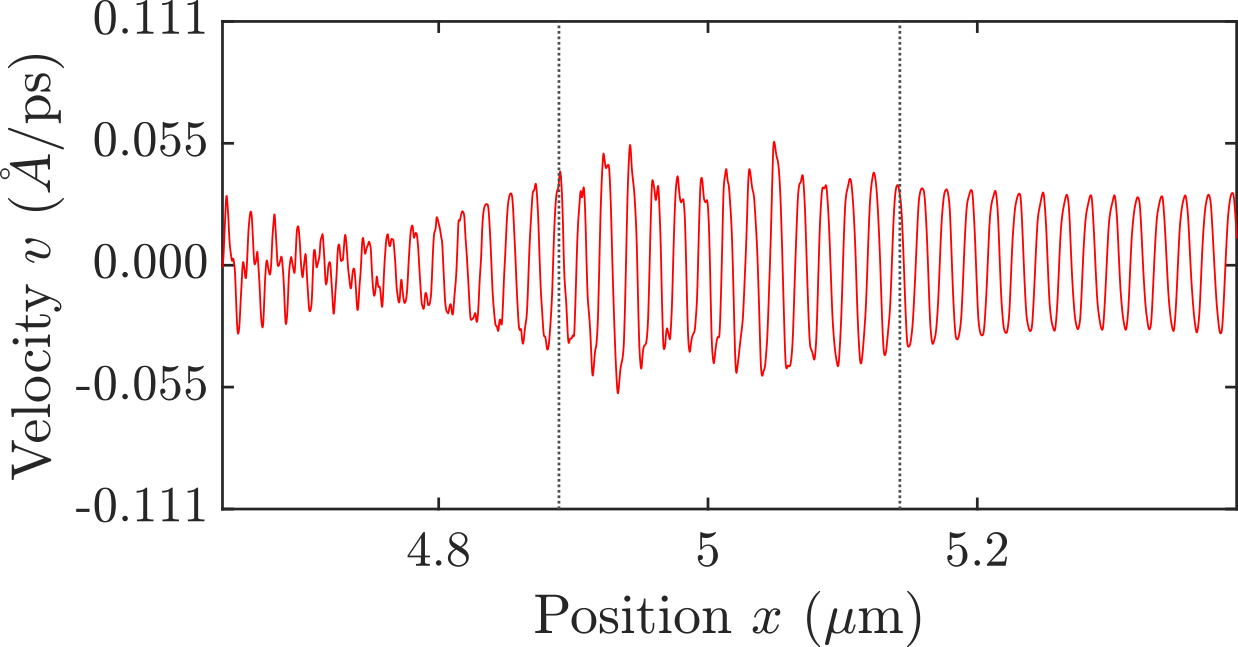}&\includegraphics[width=0.33\textwidth]{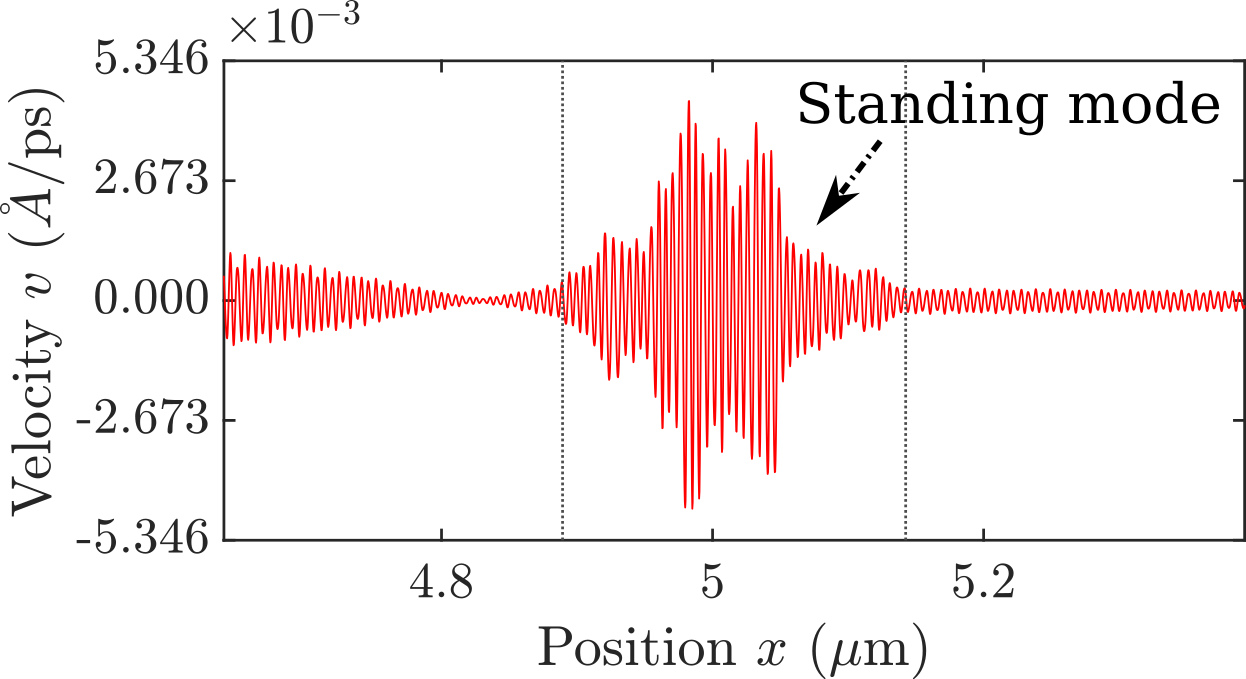} \\ \hline
\end{tabular}
\end{table*}

The amplified destructive interference leads to a reduction in the transmission of the coherent phonon, as evidenced by Fig.~\ref{fig:figspectra}. To further elucidate this phenomenon, we present snapshots (Table~\ref{tab:table1}) and accompanying movies (see Supplemental Materials) showcasing the atomic velocity profile of wave-packets characterized by three representative wavelengths (as marked in Fig.~\ref{fig:figspectra}): $\lambda_{1}$ (long), $\lambda_{2}$, and $\lambda_{4}$ (short), under the scenarios of $l_{c}=0.25L$ and $l_{c}=4L$. For long-$\lambda$ phonons, the small-$l_{c}$ wave-packet (Figs. 3a and 3b), encompassing a wide range of wavelengths in reciprocal space and thereby being less monochromatic, traverses the aperiodic SL device with notably less amplitude modulation, indicative of minimal interference. This stands in contrast to the scenarios involving large-$l_{c}$ wave-packet, as illustrated in Figs. 3d and 3e, where significant amplitude modulation is observed.

The convergence of the spectra curves, corresponding to various coherence lengths $l_{c}$, at short wavelengths $\lambda$ in Fig.~\ref{fig:figspectra}, can be attributed to pronounced destructive interference in the short-$\lambda$ limit. With identical $l_{c}$, the number of phonon wave periods in a wave-packet increases for shorter-$\lambda$ phonons, facilitating destructive interference. Moreover, shorter-$\lambda$ phonons are more susceptible to interface scattering in SL structures \cite{tamura1988acoustic,Chen_1999}. Both these factors suppress phonon transmission, thus diminishing differences in transmission at the short-$\lambda$ limit. 

Examining Figs. 3c and 3f for the short-wavelength case ($\lambda_{4}$) and the corresponding movies, it becomes evident that the initial wave-packet splits into smaller packets with a discernible phase lag between each, attributed to partial reflection or transmission at various interfaces. Noteworthy is the emergence of a standing wave within the aperiodic SL device in both Fig. 3c and Fig. 3f, a consequence of the interference between counter-propagating modes. Notably, the wave-packet with a larger $l_{c}$ (Fig. 3f) exhibits a larger amplitude, highlighting the significance of spatial coherence on interference. 

Finally, we explore the occurrence of phonon mode conversion when a coherent wave-packet interacts with the interfaces in the aperiodic SL device, considering the pronounced interference and interface scattering events revealed by velocity plots in Table ~\ref{tab:table1} and corresponding movies. The wavelet transform method allows us to directly probe reciprocal-space dynamics for different real-space positions \cite{baker2012application,chen2017effects,chen2018passing}. Here, we apply the wavelet transform to the time-dependent atomic velocities to investigate coherent phonon scattering within the aperiodic SL device. Table ~\ref{tab:table2} presents snapshots of the wavelet transform for wavelengths $\lambda_{1}$ and $\lambda_{3}$ and for coherence lengths $l_{c}=0.25L$ and $l_{c}=4L$. Full movies of these scenarios can be found in Supplemental Materials. Figs. 4a and 4b reveal that, for small-$l_{c}$ wave-packets, the reflected and transmitted components are exclusively centered around the incident wavevector $k_{0}$, indicating an absence of mode conversion. In contrast, the scattering of large-$l_{c}$ wave-packets (Figs. 4c and 4d) exhibits significant mode-conversion at long wavelengths ($\lambda_{1})$ but not at shorter wavelengths ($\lambda_{3}$). 

\begin{table*}
\tiny
\caption{\label{tab:table2}Snapshots of the reciprocal-space wavelet transform when the wave-packet scatters through the aperiodic SL device for wavelength $\lambda_{1} = 30.7$ nm and $\lambda_{3} = 9.3$ nm at spatial coherence lengths $l_{c}=0.25L$ and $l_{c}=4L$, where $L = 250$ nm. Movies of these scenarios are located in the Supplemental Materials. The dotted vertical lines denote the position and thickness of the device. The dashed horizontal line denotes the central wavevector $k_{0}$ of the incident wave-packet. Illuminated regions in the heat map not centered about the $k_{0}$ line indicates excitation of additional vibrational modes.}
\begin{tabular}{|cccc|}
\hline
\hfill&Fig. 4a: $\lambda_{1}= 30.7$ nm, $l_{c}=0.25L$, $t=424$ ps&Fig. 4b: $\lambda_{3}= 9.3$ nm, $l_{c}=0.25L$, $t=428$ ps&\hfill\\ \hline
\hfill&\includegraphics[width=0.5\textwidth]{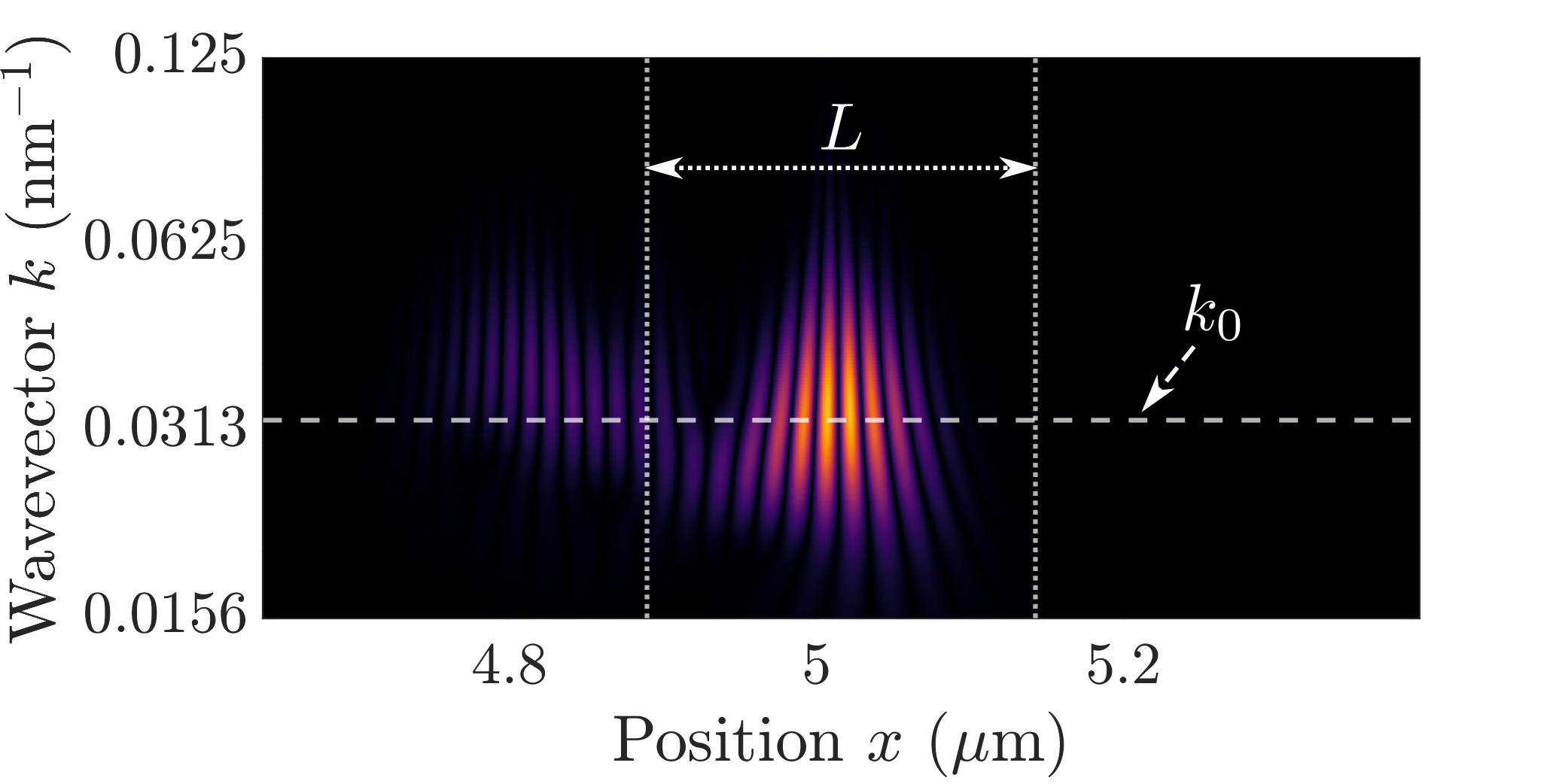}&\includegraphics[width=0.5\textwidth]{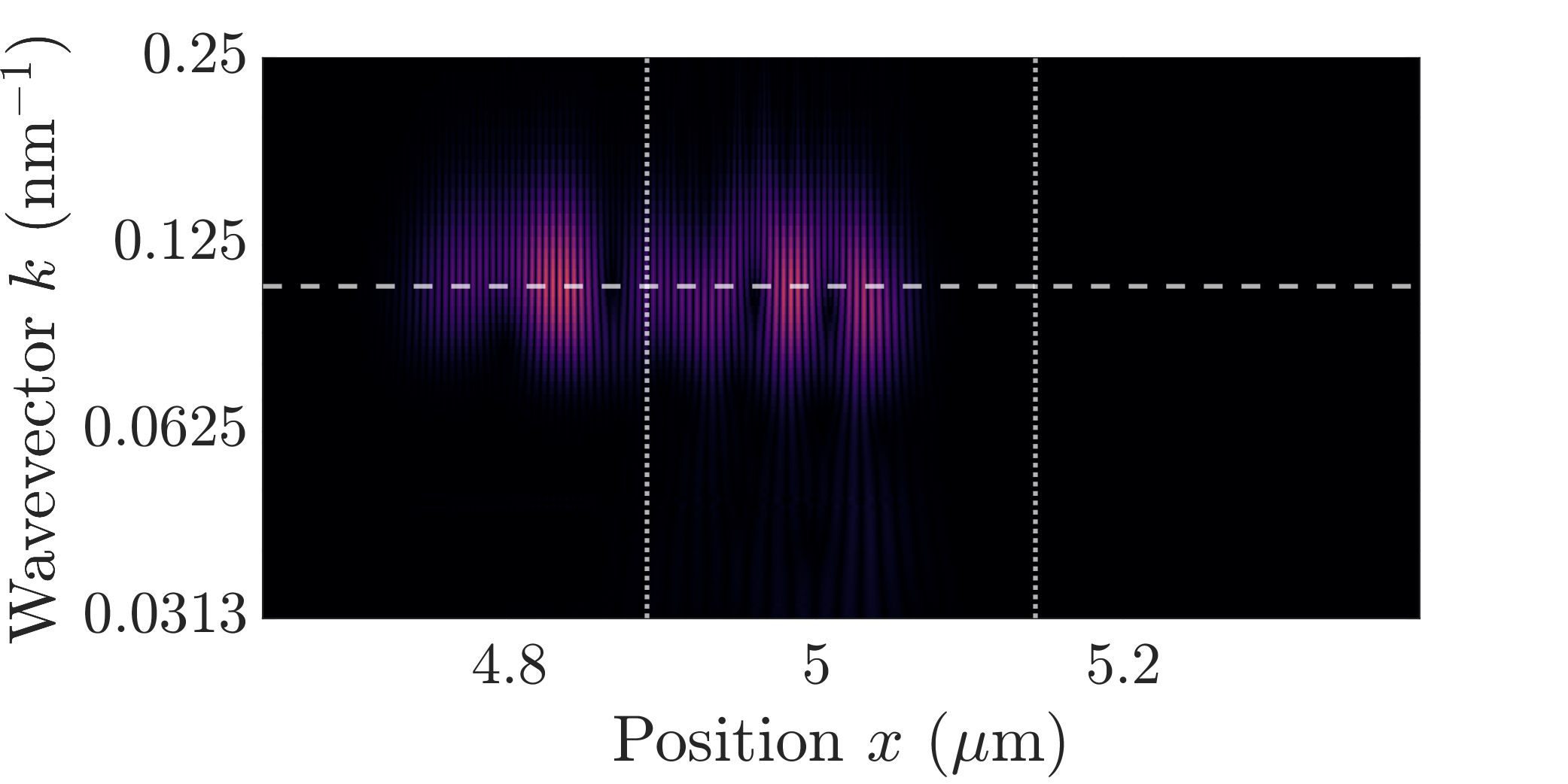}&\hfill \\ \hline
\hfill&Fig. 4c: $\lambda_{1}= 30.7$ nm, $l_{c}=4L$, $t=424$ ps&Fig. 4d: $\lambda_{3}= 9.3$ nm, $l_{c}=4L$, $t=428$ ps&\hfill\\ \hline
\hfill&\includegraphics[width=0.5\textwidth]{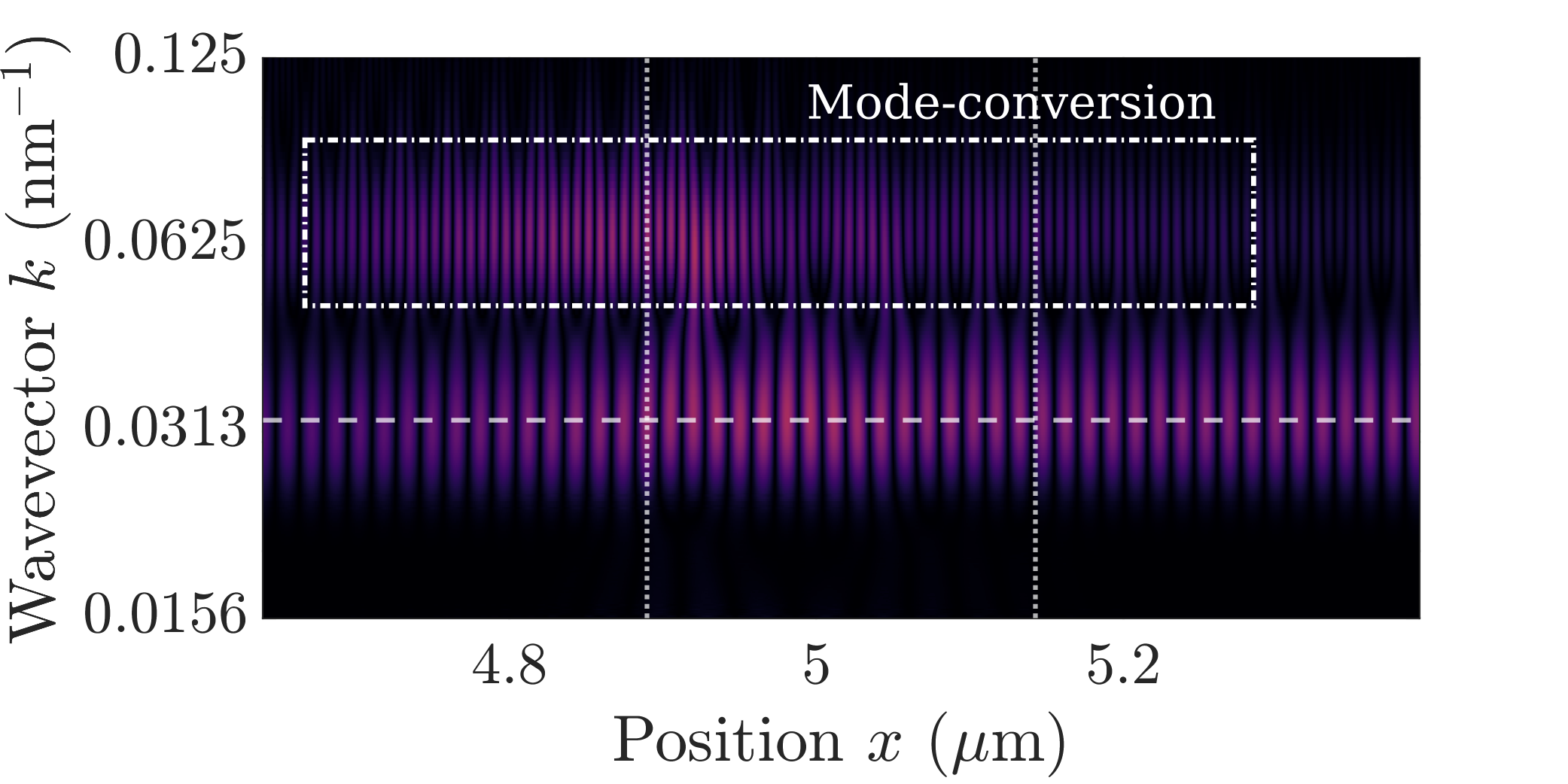}&\includegraphics[width=0.5\textwidth]{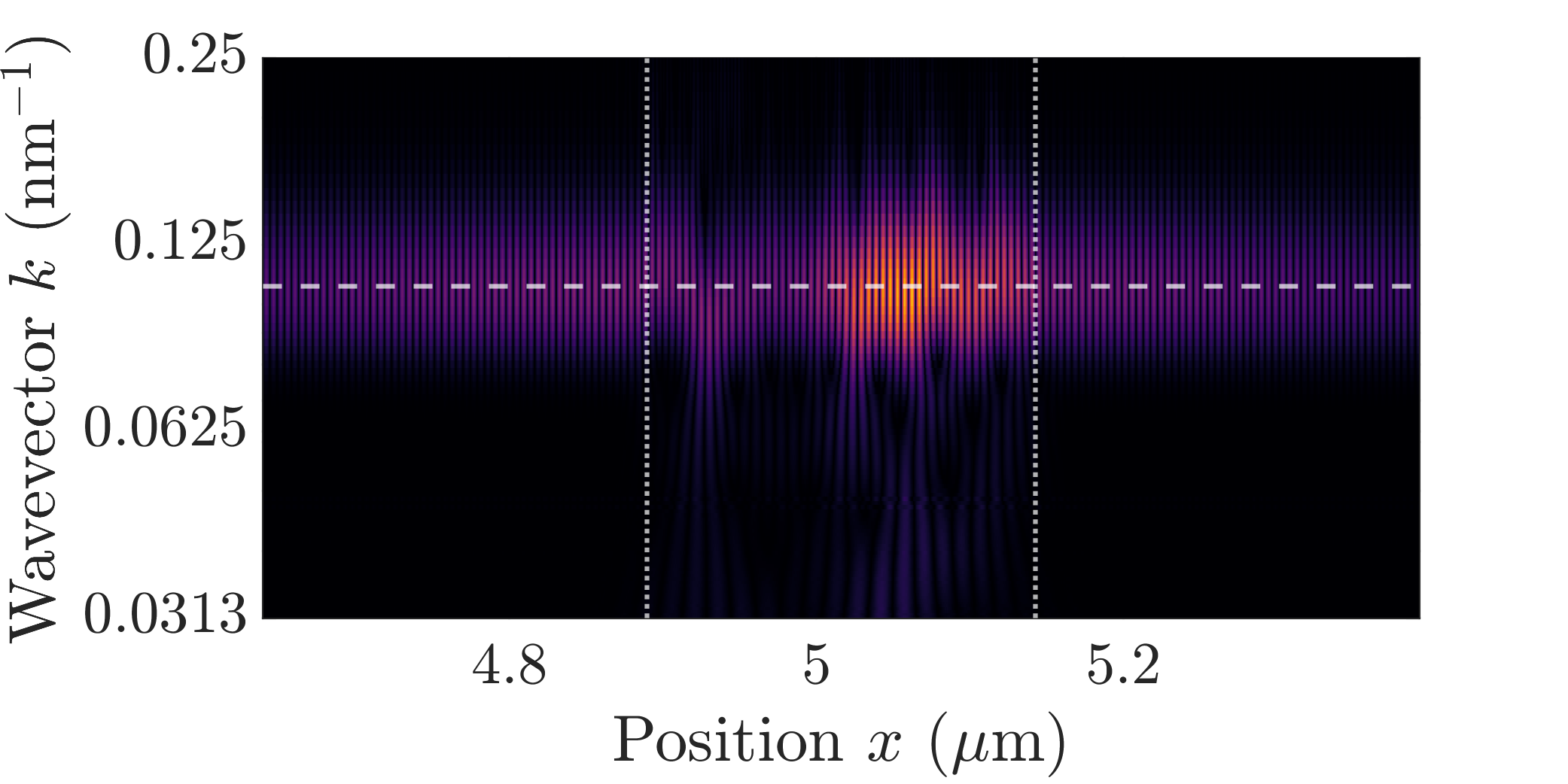}&\hfill \\ \hline
\end{tabular}
\end{table*}

In conclusion, we have investigated the propagation and scattering of coherent phonons in contrast to previously studied incoherent phonons across aperiodically arranged interfaces within the aperiodic SL structure, employing the atomistic wave-packet method. The distinctive capability of this technique to delineate spatial coherence length unveils interference dynamics in coherent phonon transmission, which depends on the relative magnitudes of phonon wavelength $\lambda$, the spatial coherence of the wave-packet $l_{c}$, and the spatial extent of the aperiodic SL device $L$. Notably, our findings demonstrate that a larger coherence length enhances destructive interference within the aperiodic SL, resulting in suppressed transmission. This introduces a novel mechanism, hitherto unreported, wherein an elevated temperature can increase the thermal conductivity of aperiodic SLs by mitigating destructive interference through the reduction of phonon coherence length. Additionally, our study reveals the excitation of new vibrational modes, i.e., mode conversion, when wave-packets with large coherence lengths traverse the aperiodic SL. This work provides valuable insights into the pivotal roles played by spatial coherence and wavelength in coherent phonon transport, advancing our comprehension of thermal transport in metamaterials or modern devices featuring a secondary periodicity. 

\section*{Acknowledgements}
The authors thank the support from the Nuclear Regulatory Commission (award number: 31310021M0004) and the National Science Foundation (award number: 2047109). The authors also would like to acknowledge the support of Research and Innovation and the Cyberinfrastructure Team in the Office of Information Technology at the University of Nevada, Reno for facilitation and access to the Pronghorn High-Performance Computing Cluster.

\section*{Conflict of Interests}
The authors have no conflicts to disclose.

\section*{Author Contributions}
\textbf{Theodore Maranets:} Conceptualization (equal), Formal analysis (equal), Methodology (lead), Software (lead), Writing—Original Draft (lead), Writing—Review and Editing (equal). \textbf{Milad Nasiri:} Conceptualization (supporting), Formal analysis (supporting), Methodology (supporting), Software (supporting), Writing—Original Draft (supporting), Writing—Review and Editing (equal). \textbf{Yan Wang:} Conceptualization (equal), Formal analysis (equal), Supervision (lead), Funding Acquisition (lead), Writing—Original Draft (supporting), Writing—Review and Editing (equal).

\section*{Data Availability Statement}
The data that support the findings of this study are available from the corresponding author upon reasonable request.

\bigskip

\bibliographystyle{unsrt}
\bibliography{references}

\begin{thebibliography}{10}

\bibitem{zhang2021coherent}
Zhongwei Zhang, Yangyu Guo, Marc Bescond, Jie Chen, Masahiro Nomura, and Sebastian Volz.
\newblock Coherent thermal transport in nano-phononic crystals: An overview.
\newblock {\em APL Materials}, 9(8), 2021.

\bibitem{anufriev2021review}
Roman Anufriev, Jeremie Maire, and Masahiro Nomura.
\newblock Review of coherent phonon and heat transport control in one-dimensional phononic crystals at nanoscale.
\newblock {\em APL Materials}, 9(7), 2021.

\bibitem{yao1987thermal}
Takafumi Yao.
\newblock Thermal properties of alas/gaas superlattices.
\newblock {\em Applied Physics Letters}, 51(22):1798--1800, 1987.

\bibitem{venkatasubramanian2000lattice}
Rama Venkatasubramanian.
\newblock Lattice thermal conductivity reduction and phonon localizationlike behavior in superlattice structures.
\newblock {\em Physical Review B}, 61(4):3091, 2000.

\bibitem{landry2008complex}
ES~Landry, MI~Hussein, and AJH McGaughey.
\newblock Complex superlattice unit cell designs for reduced thermal conductivity.
\newblock {\em Physical Review B}, 77(18):184302, 2008.

\bibitem{wang2015optimization}
Yan Wang, Chongjie Gu, and Xiulin Ruan.
\newblock Optimization of the random multilayer structure to break the random-alloy limit of thermal conductivity.
\newblock {\em Applied Physics Letters}, 106(7), 2015.

\bibitem{snyder2008complex}
G~Jeffrey Snyder and Eric~S Toberer.
\newblock Complex thermoelectric materials.
\newblock {\em Nature materials}, 7(2):105--114, 2008.

\bibitem{simkin2000minimum}
MV~Simkin and GD~Mahan.
\newblock Minimum thermal conductivity of superlattices.
\newblock {\em Physical Review Letters}, 84(5):927, 2000.

\bibitem{chen2005minimum}
Yunfei Chen, Deyu Li, Jennifer~R Lukes, Zhonghua Ni, and Minhua Chen.
\newblock Minimum superlattice thermal conductivity from molecular dynamics.
\newblock {\em Physical Review B}, 72(17):174302, 2005.

\bibitem{ravichandran2014crossover}
Jayakanth Ravichandran, Ajay~K Yadav, Ramez Cheaito, Pim~B Rossen, Arsen Soukiassian, SJ~Suresha, John~C Duda, Brian~M Foley, Che-Hui Lee, Ye~Zhu, et~al.
\newblock Crossover from incoherent to coherent phonon scattering in epitaxial oxide superlattices.
\newblock {\em Nature materials}, 13(2):168--172, 2014.

\bibitem{yang2003partially}
Bao Yang and Gang Chen.
\newblock Partially coherent phonon heat conduction in superlattices.
\newblock {\em Physical Review B}, 67(19):195311, 2003.

\bibitem{wang2014decomposition}
Yan Wang, Haoxiang Huang, and Xiulin Ruan.
\newblock Decomposition of coherent and incoherent phonon conduction in superlattices and random multilayers.
\newblock {\em Physical Review B}, 90(16):165406, 2014.

\bibitem{luckyanova2012coherent}
Maria~N Luckyanova, Jivtesh Garg, Keivan Esfarjani, Adam Jandl, Mayank~T Bulsara, Aaron~J Schmidt, Austin~J Minnich, Shuo Chen, Mildred~S Dresselhaus, Zhifeng Ren, et~al.
\newblock Coherent phonon heat conduction in superlattices.
\newblock {\em Science}, 338(6109):936--939, 2012.

\bibitem{chakraborty2020complex}
Pranay Chakraborty, Isaac~Armstrong Chiu, Tengfei Ma, and Yan Wang.
\newblock Complex temperature dependence of coherent and incoherent lattice thermal transport in superlattices.
\newblock {\em Nanotechnology}, 32(6):065401, 2020.

\bibitem{ma2022ex}
T~Ma and Y~Wang.
\newblock Ex-situ modification of lattice thermal transport through coherent and incoherent heat baths.
\newblock {\em Materials Today Physics}, 29:100884, 2022.

\bibitem{latour2014microscopic}
Benoit Latour, Sebastian Volz, and Yann Chalopin.
\newblock Microscopic description of thermal-phonon coherence: From coherent transport to diffuse interface scattering in superlattices.
\newblock {\em Physical Review B}, 90(1):014307, 2014.

\bibitem{latour2017distinguishing}
Benoit Latour and Yann Chalopin.
\newblock Distinguishing between spatial coherence and temporal coherence of phonons.
\newblock {\em Physical Review B}, 95(21):214310, 2017.

\bibitem{lee2017investigation}
Jaeho Lee, Woochul Lee, Geoff Wehmeyer, Scott Dhuey, Deirdre~L Olynick, Stefano Cabrini, Chris Dames, Jeffrey~J Urban, and Peidong Yang.
\newblock Investigation of phonon coherence and backscattering using silicon nanomeshes.
\newblock {\em Nature communications}, 8(1):14054, 2017.

\bibitem{hu2018randomness}
Shiqian Hu, Zhongwei Zhang, Pengfei Jiang, Jie Chen, Sebastian Volz, Masahiro Nomura, and Baowen Li.
\newblock Randomness-induced phonon localization in graphene heat conduction.
\newblock {\em The journal of physical chemistry letters}, 9(14):3959--3968, 2018.

\bibitem{juntunen2019anderson}
Taneli Juntunen, Osmo V{\"a}nsk{\"a}, and Ilkka Tittonen.
\newblock Anderson localization quenches thermal transport in aperiodic superlattices.
\newblock {\em Physical review letters}, 122(10):105901, 2019.

\bibitem{hu2020machine}
Run Hu, Sotaro Iwamoto, Lei Feng, Shenghong Ju, Shiqian Hu, Masato Ohnishi, Naomi Nagai, Kazuhiko Hirakawa, and Junichiro Shiomi.
\newblock Machine-learning-optimized aperiodic superlattice minimizes coherent phonon heat conduction.
\newblock {\em Physical Review X}, 10(2):021050, 2020.

\bibitem{schelling2003multiscale}
PK~Schelling and SR~Phillpot.
\newblock Multiscale simulation of phonon transport in superlattices.
\newblock {\em Journal of Applied Physics}, 93(9):5377--5387, 2003.

\bibitem{jiang2021total}
Pengfei Jiang, Yulou Ouyang, Weijun Ren, Cuiqian Yu, Jia He, and Jie Chen.
\newblock Total-transmission and total-reflection of individual phonons in phononic crystal nanostructures.
\newblock {\em APL Materials}, 9(4), 2021.

\bibitem{tamura1990acoustic}
Shin-ichiro Tamura and Franco Nori.
\newblock Acoustic interference in random superlattices.
\newblock {\em Physical Review B}, 41(11):7941, 1990.

\bibitem{bliokh2009transport}
Yury~P Bliokh, Valentin Freilikher, Sergey Savel’ev, and Franco Nori.
\newblock Transport and localization in periodic and disordered graphene superlattices.
\newblock {\em Physical Review B}, 79(7):075123, 2009.

\bibitem{roy2022unexpected}
Prabudhya Roy~Chowdhury and Xiulin Ruan.
\newblock Unexpected thermal conductivity enhancement in aperiodic superlattices discovered using active machine learning.
\newblock {\em npj Computational Materials}, 8(1):12, 2022.

\bibitem{schelling2002phonon}
PK~Schelling, SR~Phillpot, and P~Keblinski.
\newblock Phonon wave-packet dynamics at semiconductor interfaces by molecular-dynamics simulation.
\newblock {\em Applied Physics Letters}, 80(14):2484--2486, 2002.

\bibitem{jones1924determination}
John~Edward Jones.
\newblock On the determination of molecular fields.—ii. from the equation of state of a gas.
\newblock {\em Proceedings of the Royal Society of London. Series A, Containing Papers of a Mathematical and Physical Character}, 106(738):463--477, 1924.

\bibitem{huberman2013disruption}
Samuel~C Huberman, Jason~M Larkin, Alan~JH McGaughey, and Cristina~H Amon.
\newblock Disruption of superlattice phonons by interfacial mixing.
\newblock {\em Physical Review B}, 88(15):155311, 2013.

\bibitem{chakraborty2020quenching}
Pranay Chakraborty, Yida Liu, Tengfei Ma, Xixi Guo, Lei Cao, Run Hu, and Yan Wang.
\newblock Quenching thermal transport in aperiodic superlattices: A molecular dynamics and machine learning study.
\newblock {\em ACS applied materials \& interfaces}, 12(7):8795--8804, 2020.

\bibitem{thompson2022lammps}
Aidan~P Thompson, H~Metin Aktulga, Richard Berger, Dan~S Bolintineanu, W~Michael Brown, Paul~S Crozier, Pieter~J in't Veld, Axel Kohlmeyer, Stan~G Moore, Trung~Dac Nguyen, et~al.
\newblock Lammps-a flexible simulation tool for particle-based materials modeling at the atomic, meso, and continuum scales.
\newblock {\em Computer Physics Communications}, 271:108171, 2022.

\bibitem{hopkins2009multiple}
Patrick~E Hopkins.
\newblock Multiple phonon processes contributing to inelastic scattering during thermal boundary conductance at solid interfaces.
\newblock {\em Journal of Applied Physics}, 106(1), 2009.

\bibitem{tamura1988acoustic}
SI~Tamura, DC~Hurley, and JP~Wolfe.
\newblock Acoustic-phonon propagation in superlattices.
\newblock {\em Physical Review B}, 38(2):1427, 1988.

\bibitem{Chen_1999}
G.~Chen.
\newblock Phonon wave heat conduction in thin films and superlattices.
\newblock {\em Journal of Heat Transfer}, 121(4):945–953, Nov 1999.

\bibitem{baker2012application}
Christopher~H Baker, Donald~A Jordan, and Pamela~M Norris.
\newblock Application of the wavelet transform to nanoscale thermal transport.
\newblock {\em Physical Review B}, 86(10):104306, 2012.

\bibitem{chen2017effects}
Xiang Chen, Liming Xiong, David~L McDowell, and Youping Chen.
\newblock Effects of phonons on mobility of dislocations and dislocation arrays.
\newblock {\em Scripta Materialia}, 137:22--26, 2017.

\bibitem{chen2018passing}
Xiang Chen, Adrian Diaz, Liming Xiong, David~L McDowell, and Youping Chen.
\newblock Passing waves from atomistic to continuum.
\newblock {\em Journal of Computational Physics}, 354:393--402, 2018.

\end{thebibliography}
\end{document}